\documentclass[useAMS]{article}
\usepackage{graphicx}
\usepackage{txfonts}
\usepackage[square,comma,numbers]{natbib}

\def\beq#1{\begin{equation}\label{#1}}
\def\eeq{\end{equation}}
\def\beqa#1{\begin{eqnarray}\label{#1}}
\def\eeqa{\end{eqnarray}}

\def\Eq#1{Eq.~(\ref{#1})}

\def\myfrac#1#2{\left(\frac{#1}{#2}\right)}

\def\mycomment#1{\relax}

\begin{document}

\title{Wind accretion: Theory and Observations}
\author{N.I. Shakura$^1$, K.A. Postnov$^1$,\\
 A.Yu. Kochetkova$^1$, L. Hjalmarsdotter$^1$,\\
 L. Sidoli$^2$, A. Paizis$^2$
 \\
\small\itshape 
$^1$ Moscow M.V. Lomonosov State University,\\
\small\itshape  Sternberg Astronomical Institute,
13, Universitetskij pr., 119992 Moscow, Russia\\
\small\itshape $^{2}$ INAF, Istituto di Astrofisica Spaziale e Fisica Cosmica, \\
\small\itshape Via E.\ Bassini 15,   I-20133 Milano,  Italy
} 
\maketitle

\begin{abstract}
A review of wind accretion in high-mass X-ray binaries is presented. We focus on 
different regimes of quasi-spherical accretion onto a neutron star: supersonic (Bondi)
accretion, which takes place when the captured matter cools down rapidly and falls
supersonically towards the neutron-star 
magnetosphere, and subsonic (settling) accretion which occurs when
the plasma remains hot until it meets the magnetospheric boundary. The two regimes of accretion are separated by a limit in X-ray luminosity at about 
$4\times 10^{36}$~erg~s$^{-1}$. In subsonic accretion, which works at
lower luminosities, a hot quasi-spherical shell must form around the magnetosphere, and the
actual accretion rate onto the neutron star is determined by the ability of the plasma to enter the
magnetosphere due to the Rayleigh-Taylor instability. In turn, two regimes of subsonic accretion
are possible, depending on the plasma cooling mechanism (Compton or radiative) near the
magnetopshere. The transition from the high-luminosity regime  
with Compton cooling to the low-luminosity ($L_x \lesssim 3\times 10^{35}$~erg~s$^{-1}$) regime with 
radiative cooling can be responsible for the onset of
the 'off« states repeatedly observed in several low-luminosity slowly accreting pulsars, such as
Vela X-1, GX 301-2 and 4U 1907+09. The triggering of the transition may be due to a switch
in the X-ray beam pattern in response to a change in the optical depth in the accretion column
with changing luminosity. 
We also show that in the settling accretion theory, bright X-ray flares ($\sim 10^{38}-10^{40}$~ergs) observed in supergiant fast X-ray transients (SFXT)  may be produced by 
sporadic capture of magnetized stellar-wind plasma. At sufficiently low accretion rates, 
magnetic reconnection 
can enhance the magnetospheric plasma entry rate, resulting in copious production of X-ray photons,
strong Compton cooling and ultimately in unstable 
accretion of the entire shell. 
A bright flare develops on the free-fall time
scale in the shell, and the typical energy released in an SFXT bright flare corresponds to 
the mass of the shell.

\end{abstract}

\section{Introduction}
\label{intro}
In 1966, one of us, Nikolay Ivanovich Shakura, was a 4th-year student at the 
Physical Department of Moscow State University,
and heard for the first time the term 'accretion' from Yakov Borisovich Zeldovich. 
Yakov Borisovich suggested to Nikolay Ivanovich to study spherical accretion onto a neutron star (NS). Note 
that at that time neither radio pulsars, nor X-ray pulsars or accreting black holes
were known -- all these rich observational {\bf appearances} of neutron stars and black holes were 
discovered later. When formulating this problem, Zeldovich was motivated by the discovery of galactic X-ray 
sources from rocket flights, such as Sco X-1. 
\citep{1962PhRvL...9..439G} (see the Nobel lecture by R. Giacconi 
\citep{2003RvMP...75..995G} for an historical review). 

During spherical accretion of 
gas onto a NS without magnetic field, a strong shock arises above the NS surface, 
the gas is heated up 
to a temperature of up to a few keV-. The structure of this region and 
the emergent radiation spectrum was first calculated in \citep{1969SvA....13..175Z}. 
While the mechanism of X-ray emission from a spherically accreting NS as proposed in that paper 
reproduced the main features of the observed X-ray spectra,
the nature of galactic X-ray sources remained unclear until the launch of the first specialized UHURU satellite. UHURU discovered two main types of X-ray sources -- those which show 
regular pulsations like Cen X-3 and Her X-1 (X-ray pulsars), and those with chaotic 
X-ray variability like Cyg X-1 (which turned out to be a black-hole candidate). Both are cases of
accretion in a close binary system in which matter is transferred from the optical 
(non-degenerate) star overfilling its Roche lobe onto the compact object (a degenerate star or a black hole) with the formation of an accretion disk \citep{1971ApJ...167L..67G, 1973IAUS...55.....B}. \footnote{Note that between the first discoveries of Galactic X-ray sources in 1962 and 
the UHURU observations, 
radio pulsars with non-trivial (and unclear up to now!) 
non-thermal mechanisms of radio emission were reported \citep{1968Natur.217..709H}. Soon they 
were recognized to be isolated rotating magnetized NS \citep{1968Natur.218..731G, 
1969ApJ...157..869G}}. 

\section{Supersonic and subsonic wind accretion}

Generally, in close binary systems, 
there can be two different regimes of accretion onto the compact object -- disk accretion 
\citep{1973SvA....16..756S,1972A&A....21....1P,1973A&A....24..337S}\footnote{Accretion disks in
active galactic nuclei were first considered by Lynden-Bell \citep{1969Natur.223..690L}.}
and quasi-spherical accretion. The disk accretion regime is usually realized 
when the optical star overfills its Roche lobe. Quasi-spherical accretion is 
most likely to occur in high-mass X-ray binaries (HMXB) 
when an optical star of early spectral class (O-B) does not fill its Roche lobe, 
but experiences a significant mass loss via its stellar wind. We shall discuss the wind accretion regime,
in which a bow shock forms in the stellar wind around the compact star. The structure of the bow 
shock and the associated accretion wake is non-stationary and quite complicated (see e.g.
numerical simulations \citep{1988ApJ...335..862F, 1999A&A...346..861R, 2004A&A...419..335N}, 
among many others). The characteristic distance at which the bow shock forms
is approximately that of the Bondi radius $R_B=2GM/(v_w^2+v_{orb}^2)$, where $v_w$ is the wind velocity 
(typically 100-1000 km/s) and $v_{orb}$ is the orbital velocity of the compact star. In HMXBs, the stellar wind velocity is usually much larger than $v_{orb}$, so below we will neglect $v_{orb}$. 
The rate of gravitational capture of mass from a wind with density $\rho_w$
near the orbital position of the NS  
is the Bondi mass accretion rate: $\dot M_B\simeq \rho_w R_B^2 v_w\propto \rho_w v_w^{-3}$.   

\begin{figure*}
\includegraphics[width=7cm]{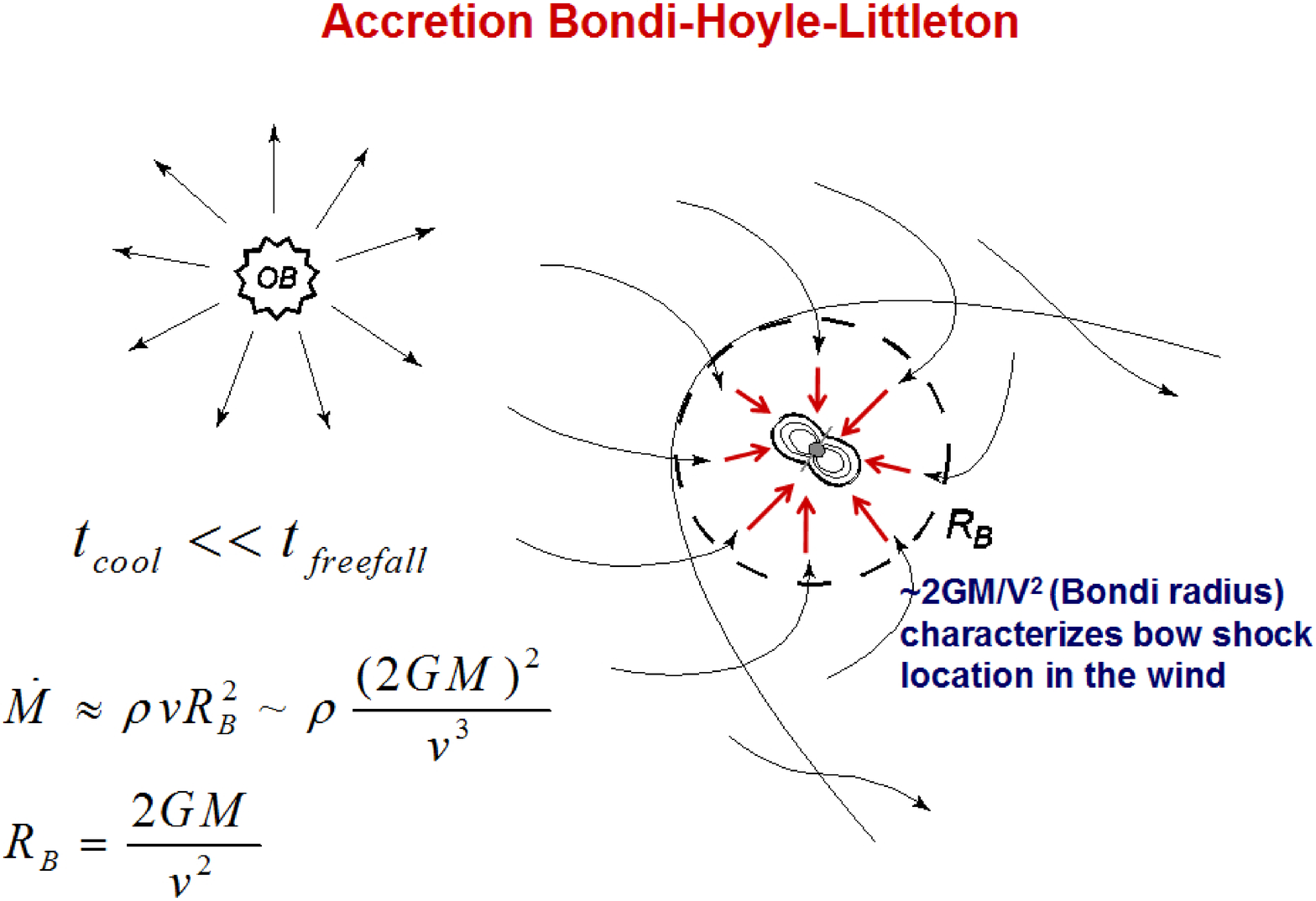}
\includegraphics[width=7cm]{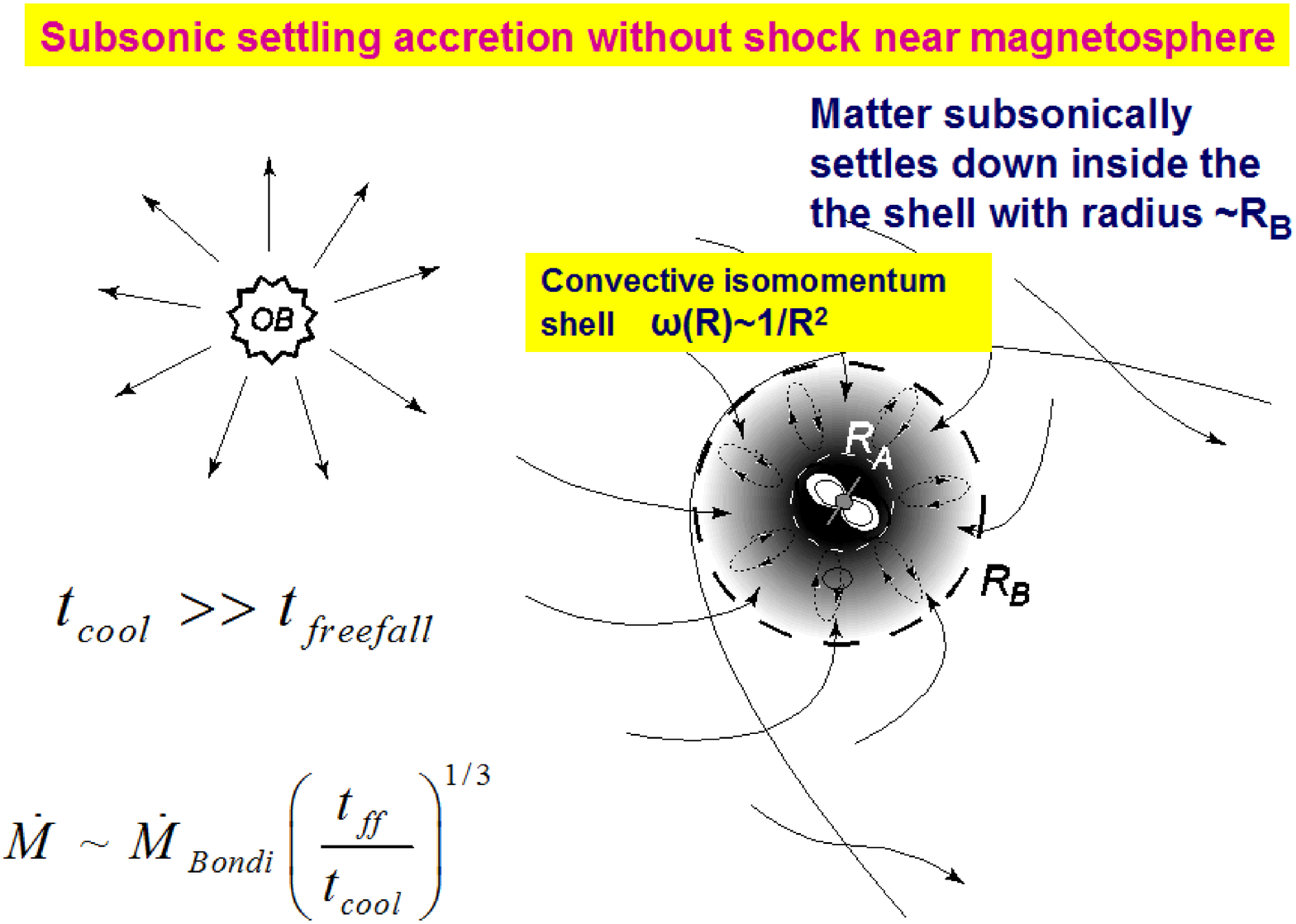}
\parbox[t]{0.47\textwidth}{\caption{Supersonic (Bondi-Hoyle-Littleton) accretion onto magnetized NS}\label{f:1}}
\hfill
\parbox[t]{0.47\textwidth}
{\caption{Subsonic settling accretion onto magnetized NS}\label{f:2}}
\end{figure*}

Then, there are two different cases of quasi-spherical accretion. Classical 
Bondi-Hoyle-Littleton 
accretion takes place when the shocked matter is cooled down rapidly, 
and the matter falls freely towards the NS magnetosphere
(see Fig. \ref{f:1}) by forming 
a shock at some distance above the magnetosphere. 
Here the shocked matter cools down (mainly via Compton processes) 
and enters the magnetopshere due to the Rayleigh-Taylor instability \citep{1976ApJ...207..914A}.
The magnetospheric boundary is characterized by the Alfv\'en radius $R_A$, which can be
calculated from the balance between the ram pressure of the 
infalling matter and the magnetic field pressure at the 
magnetospheric boundary. 
The captured matter from the wind carries a specific angular momentum 
$j_w\sim \omega_BR_B^2$ \citep{1975A&A....39..185I}. 
Depending on the sign of $j_w$ (prograde or retorgrade), the NS can spin-up 
or spin-down. This regime of quasi-spherical accretion occurs in 
bright X-ray pulsars with $L_x>4\times 10^{36}$~erg~s$^{-1}$ \citep{1983ApJ...266..175B, 2012MNRAS.420..216S}. 

If the captured wind matter behind the bow shock at $R_B$ remains hot (which it does when 
the plasma cooling time is much longer than the free-fall time, $t_{cool}\gg t_{ff}$), 
a hot quasi-static shell forms around the magnetosphere and subsonic 
(settling) accretion sets in (see Fig. \ref{f:2}).  
In this case, both spin-up and spin-down of the NS 
is possible, even if the sign of $j_w$ is positive (prograde). The shell mediates the
angular momentum transfer from the NS magnetosphere via viscous stresses
due to convection and turbulence. In this regime, the mean radial velocity 
of matter in the shell $u_r$ is smaller than the free-fall velocity $u_{ff}$: 
$u_r=f(u)u_{ff}$, $f(u)<1$, and is determined by the palsma cooling rate 
near the magnetosphere (due to Compton or radiative cooling): 
$f(u)\sim [t_{ff}(R_A)/t_{cool}(R_A)]^{1/3}$. In the settling accretion regime 
the actual mass accretion rate
onto the NS may be significantly smaller than the Bondi mass accretion rate, $\dot M=f(u) \dot M_B$.  
Settling accretion occurs 
at $L_x<4\times 10^{36}$~erg~s$^{-1}$ \citep{2012MNRAS.420..216S}. 

\section{Two regimes of plasma entering the NS magnetosphere}

\begin{figure*}
\includegraphics[width=0.7\textwidth]{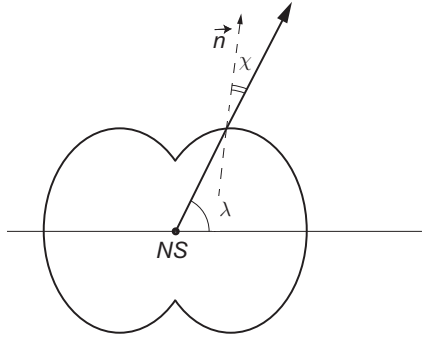}
\caption{Schematics of the NS magnetosphere}
\label{f:nsangles}
\end{figure*}

To enter the magnetosphere, the plasma in the shell must cool down from a  
high (almost virial) temperature $T$ determined by hydrostatic equilibrium 
to some critical temperature $T_{cr}$ \citep{1977ApJ...215..897E} 
\beq{30}
{\cal R}T_{cr}=\frac{1}{2}\frac{\cos\chi}{\kappa R_A}\frac{\mu_mGM}{R_A}
\eeq
Here ${\cal R}$ is the universal gas constant, $\mu_m\approx 0.6$ is the molecular weight,
$G$ is the Newtonian gravitational constant, $M$ is the neutron star mass,  
$\kappa$ is the local curvature of the magnetosphere and $\chi$ is the angle 
between the outer normal and the radius-vector at any given point at the Alfv\'en surface
(Fig. \ref{f:nsangles}).

As was shown in \citep{2012MNRAS.420..216S, 2014EPJWC..6402001S}, a transition zone above the 
Alfv\'en surface with radius $R_A$ is formed inside which the plasma cools down. The effective gravitational acceleration in this zone is 
\beq{}
g_{eff}=\frac{GM}{R_A^2}\cos\chi \left(1-\frac{T}{T_{cr}}\right)
\eeq 
and the mean radial velocity 
of plasma settling is 
\beq{}
u_R=f(u)\sqrt{2GM/R_A}\,.
\eeq
In the steady state, the dimensionless factor  
$0\le f(u)\le 1$ is determined by the specific plasma cooling mechanism 
in this zone and, by conservation of mass, is constant through the shell.  
This factor can be expressed through 
the plasma cooling time $t_{cool}$ in the transition zone \citep{2012MNRAS.420..216S}:
\beq{fu}
f(u) \simeq \myfrac{t_{ff}}{t_{cool}}^{1/3}\cos\chi^{1/3}
\eeq
where $t_{ff}=R^{3/2}/\sqrt{2GM}$ is the characteristic free-fall time scale from radius $R$. The angle
$\chi$ is determined by the shape of the magnetosphere, and for the magnetospheric boundary 
parametrized in the form   
$\sim \cos\lambda^n$ (where $\lambda$ is the angle calculated from the magnetospheric equator, see Fig. 
\ref{f:nsangles}) 
$\tan \chi=n\tan\lambda$. For example, in model calculations by \citep{1976ApJ...207..914A} $n\simeq 0.27$ in the near-equatorial zone, so $\kappa R_A\approx 1.27$.
We see that $\cos\chi\simeq 1$ up to $\lambda\sim \pi/2$, 
so below (as in \citep{2012MNRAS.420..216S}) we shall omit $\cos\chi$. 

Along with the density of 
matter near the magnetospheric boundary $\rho(R_A)$, 
the factor $f(u)$ determines the magnetosphere 
mass loading rate through 
the mass continuity equation:
\beq{cont}
\dot M=4\pi R_A^2 \rho(R_A) f(u)\sqrt{2GM/R_A}\,.
\eeq
This plasma eventually reaches the neutron star surface and 
produces an X-ray luminosity $L_x\approx 0.1\dot M c^2$.
Below we shall normalize the mass accretion rate 
through the magnetosphere as well as the X-ray luminosity 
to the fiducial values $\dot M_{n}\equiv \dot M/10^{n}$~g~s$^{-1}$ 
and $L_{n}\equiv L_x/10^{n}$~erg~s$^{-1}$, respectively.

\subsection{The Compton cooling regime}

As explained in detail in \citep{2012MNRAS.420..216S} (Appendix C and D), in subsonic quasi-static shells above slowly rotating NS magnetospheres 
the adiabaticity of the accreting matter is broken due to turbulent heating and Compton cooling. 
X-ray photons generated near the NS surface tend to cool down the matter in the shell via Compton scattering as long as the plasma
temperature $T>T_x$, where $T_x$ is the characteristic radiation temperature determined
by the spectral energy distribution of the X-ray radiation. 
For typical X-ray pulsars $T_x\sim 3-5$~keV. 
Cooling of the plasma at the base of the shell decreases the temperature gradient 
and hampers convective motions. Additional heating due to   
sheared convective motions is insignificant (see Appendix C of \citep{2012MNRAS.420..216S}).  
Therefore, the temperature in the shell 
changes with radius almost
adiabatically ${\cal R}T\sim (2/5)GM/R$, and 
the distance $R_x$ within which the plasma cools down 
by Compton scattering is
\beq{}
R_x\approx 10^{10}\hbox{cm}\myfrac{T_x}{3\hbox{keV}}^{-1}\,, 
\eeq,
much larger than the characteristic Alfv\'en radius $R_A\simeq 10^9$~cm.

The Compton cooling time is inversely proportional to the 
photon energy density, 
\beq{t_C}
t_C\sim R^2/L_x\,, 
\eeq 
and near the Alfv\'en surface we find
\beq{t_Cn}
t_C\approx 10 \hbox{[s]} \myfrac{R_A}{10^9 \hbox{cm}}^2 
L_{36}^{-1}\,.
\eeq
(This estimate assumes spherical symmetry of the X-ray emission beam). 
Clearly, for the exact radiation density the shape of the X-ray emission  
produced in the accretion column near the NS surface (i.e. X-ray beam) 
is important, but still $L_x\sim \dot M$. 
Therefore, roughly, $f(u)_C\sim \dot M^{1/3}$, or, more precisely, 
taking into account the dependence of $R_A$ on $\dot M$, in this regime
\beq{RA_C}
R_A^C\approx 10^9\hbox{cm}L_{36}^{-2/11}\mu_{30}^{6/11}
\eeq
we obtain:
\beq{fu_C}
f(u)_C\approx 0.3L_{36}^{4/11}\mu_{30}^{-1/11}\,.
\eeq
Here $\mu_{30}=\mu/10^{30}$~G cm$^3$ is the NS dipole magnetic moment.

\subsection{The radiative cooling regime}

In the absence of a dense photon field, 
at the characteristic temperatures near the magnetosphere $T\sim 30$-keV and higher,
plasma cooling is essentially due to radiative losses (bremsstrahlung), and the plasma cooling time is 
$t_{rad}\sim \sqrt{T}/\rho$. Making use of the continuity equation (\ref{cont})
and the temperature distribution in the shell $T\sim 1/R$, we obtain 
\beq{t_rad}
t_{rad}\sim R \dot M^{-1} f(u)\,.
\eeq
Note that, unlike the Compton cooling time (\ref{t_C}), the radiative cooling time is 
actually independent of $\dot M$ (remember that $\dot M\sim f(u)$ in the subsonic accretion
regime!). 
Numerically, near the magnetosphere we have 
\beq{t_radn}
t_{rad}\approx 1000 \hbox{[s]} \myfrac{R_A}{10^9 \hbox{cm}}L_{36}^{-1}\myfrac{f(u)}{0.3}\,.
\eeq

Following the method described in Section 3 of \cite{2012MNRAS.420..216S}, we find the 
mean radial velocity of matter entering the NS magnetosphere in the 
near-equatorial region,
similar to the expression for $f(u)$ in the Compton cooling region \Eq{fu_C}. 
Using the expression for the Alfv\'en radius as expressed through $f(u)$, we calculate
the dimensionless settling velocity:  
\beq{fu_rad1}
f(u)_{rad}\approx 0.1 L_{36}^{2/9}\mu_{30}^{2/27}
\eeq 
and the Alfv\'en radius:
\beq{RA_rad}
R_A^{rad}\simeq 10^9\hbox{[cm]} L_{36}^{-2/9}\mu_{30}^{16/27}
\eeq
(in the numerical estimates we assume a monoatomic gas with adiabatic index 
$\gamma=5/3$). 
The obtained expression for the dimensionless settling velocity of matter 
\Eq{fu_rad1} in the radiative cooling regime clearly shows that here accretion proceeds
much less effectively than in the Compton cooling regime (cf. with \Eq{fu_C}).   

Unlike in the Compton cooling regime, 
in the radiative cooling regime there is no instability leading to an increase of the mass
accretion rate as the luminosity increases (due to the long characteristic
cooling time), 
and accretion here is therefore expected to proceed more quietly
under the same external conditions.


\section{Comparison with observations}
\subsection{High and low ('off') states in X-ray pulsars}

So called `off' states have been observed in several slowly rotating low luminosity pulsars such as Vela X-1 \citep{Inoue_ea84, Kreykenbohm_ea99, Kreykenbohm_ea08, Doroshenko_ea11}, 
GX 301-2 \citep{Gogus_ea11} and 4U 1907+09 \citep{intZand_ea97, Sahiner_ea12, 2012A&A...548A..19D}). 
These states are characterized by a sudden, most often without any prior indication, drop in X-ray flux down to $1-10 \%$ of normal levels, lasting typically for a few minutes. 


It seems to be fairly well established that the off states can not be simply due to increased absorption along the line of sight. Their occurrence is not correlated with increased $N_H$ 
(\citep{Furst_ea11},\citep{Sahiner_ea12} but see also\citep{Kretschmar_ea99} for a another type of intensity dips in Vela X-1, most probably caused by dense blobs in the wind), the timescale of their onsets is too short (e.g.\citep{Kreykenbohm_ea08})
and spectral studies show a softening of the X-ray spectrum during the off state \citep{Gogus_ea11, Doroshenko_ea11}, contrary to what expected had the decreased flux levels been caused by increased absorption. 
Failure by early observations with instruments like \textit{RXTE}/PCA to detect pulsations during the 
off states have seemed to suggest that the sources were instead simply turned off due to 
a sudden cessation of accretion. The popular view is that the cause of this may be large 
density variations in the stellar wind, possibly combined with the onset of the propeller regime (see e.g.\citep{Kreykenbohm_ea08}). 

Recent observations with the more sensitive instruments onboard Suzaku of Vela X-1  
\citep{Doroshenko_ea11} and 4U 1907+09 \citep{2012A&A...548A..19D}, however, show that although dropping in luminosity by a factor of about 20 
the sources are clearly detected with pulse periods equal to those observed at normal flux levels. 
This suggests that rather than cessation of accretion, the off-states may be better explained 
by a transition to a different, less effective, accretion regime. We suggest that the onset 
of the off state in these sources marks a transition from the Compton cooling dominated 
to the radiative cooling dominated regime when the accretion rate drops below some value 
$L_\dag$. 

\subsection{Switch between pencil and fan beam from the accretion column}

\begin{figure*}
\includegraphics[width=0.9\textwidth]{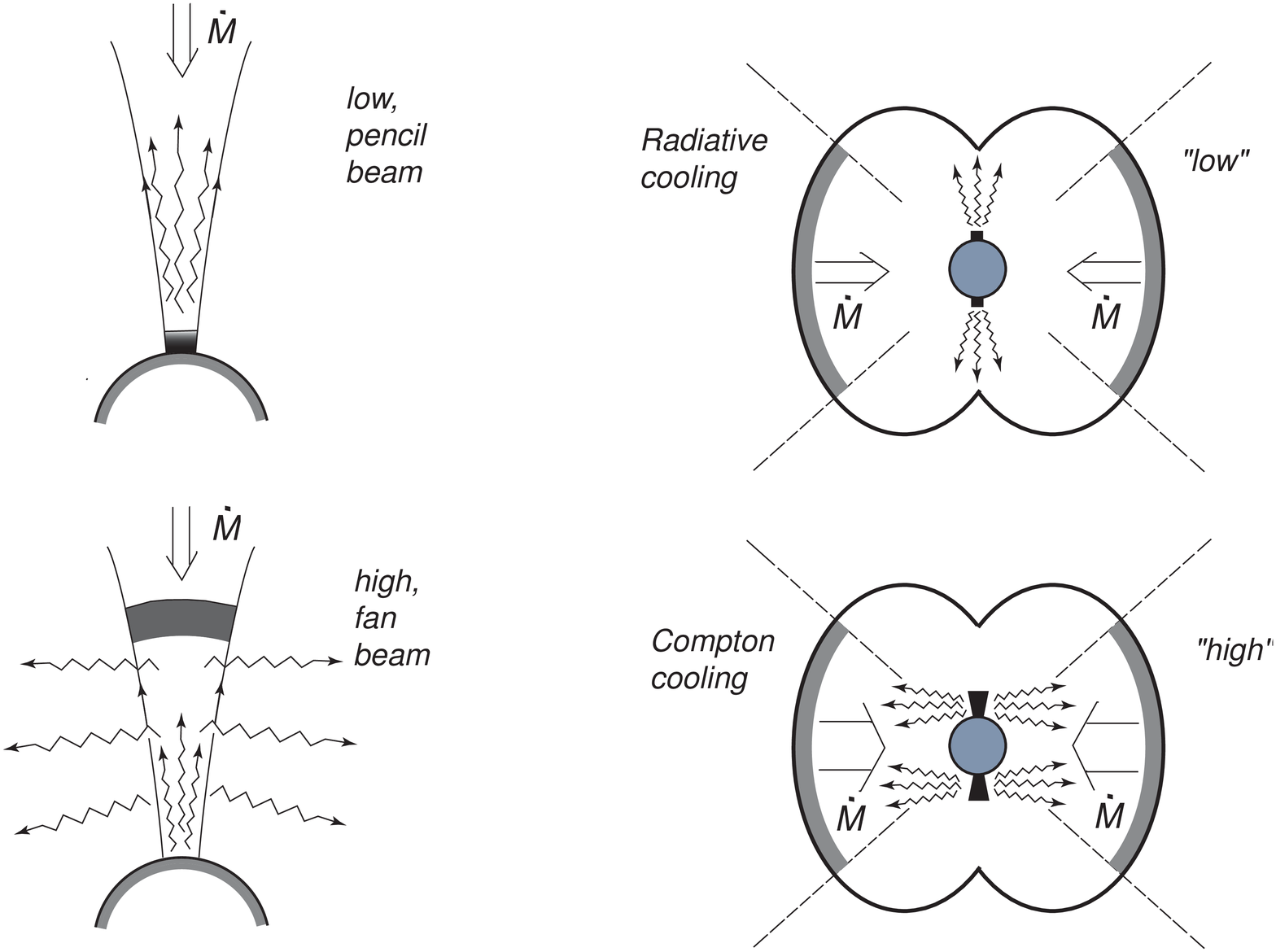}
\caption{Schematics of the transition between the Compton ('high' state) and
radiative ('low' state) plasma cooling regimes due to the X-ray beam switch
from fan to pencil pattern with decreasing mass accretion rate.}
\label{f:switch}
\end{figure*}

A decrease in the X-ray photon energy density in the transition zone diminishes
the Compton cooling efficiency, but the Compton cooling time remains much
shorter than the radiative cooling time down to very low luminosities
(see \Eq{t_Cn} and \Eq{t_radn}). Therefore, in the spherically symmetric case, a transition between the two regimes would
require an almost complete switch-off of Compton cooling in the
equatorial magnetospheric region. In the more realistic 
non-spherical case, the Compton cooling time can become comparable 
to the radiation cooling time when the X-ray beam pattern changes 
with decreasing X-ray luminosity from a fan beam to a pencil beam, and the 
equatorial X-ray flux is reduced by a factor of a few. Additionally, hardening of the pulsed X-ray 
flux with decreasing X-ray luminosity, which is observed in low-luminosity X-ray pulsars \citep{Klochkov_ea11}, increases $T_x$ and decreases the specific Compton cooling rate of the
plasma $\propto (T-T_x)/t_C$, thus making Compton cooling less efficient. 
Such X-ray pattern transitions have been observed in transient X-ray pulsars 
(see, e.g., \citep{Parmar_ea89}).
The radiation density in the X-ray pencil beam
cools down the plasma predominantly in the magnetospheric cusp region, but because of 
the stronger magnetic line curvature \citep{1976ApJ...207..914A} the plasma entry rate  
through the cusp will be insignificant. Still, the plasma 
continues to enter the magnetosphere via instabilities 
in the equatorial zone, but at 
a lower rate determined by the longer radiative cooling timescale.  
This is illustrated in Fig. \ref{f:switch}.

The mass accretion rate in the radiative cooling regime 
will be determined by the plasma density by the time 
Compton cooling switches off in the magnetospheric equator region.
This occurs at some X-ray luminosity $L_x\lesssim L_{\dag}$, which 
can be estimated from the following considerations. 

The transition from fan to pencil beam in low-luminosity X-ray 
pulsars with $L_x<10^{37}$~erg~s$^{-1}$, where no high accretion
columns should exist, does not occur until the optical depth in the accretion flow
above the polar cap becomes less than one. 
The optical depth in the accretion flow in the direction 
normal to the NS surface from the radial distance $r_6=r/10^6$~cm 
is estimated to be \citep{Lamb_ea73}
\beq{tau_n}
\tau_n\simeq 3 \myfrac{R_A}{10^9\hbox{cm}}^{1/2}\dot M_{16} r_6^{-3/2}
\eeq
(here the NS mass is assumed to be 1.5 $M_\odot$ and the NS radius $R_{NS}=10^6$~cm).
Taking into account the dependence of the Alfv\'en radius on $\dot M$ and $\mu$ (see Eq. (\ref{RA_C})), 
we see from this estimate that the X-ray emission beam change is expected 
to occur at $\tau_n<1$, corresponding to an X-ray luminosity of  
\beq{L_cross}
L_{\dag}\sim 3\times 10^{35}[\hbox{erg~s$^{-1}$}]\mu_{30}^{-3/10}\,. 
\eeq

The return from radiative cooling dominated accretion back to the Compton cooling
dominated regime can take place, for example, due to 
a density increase above the magnetosphere, leading to an increase in the  
mass accretion rate. The higher accretion rate leads to 
growth of the vertical optical depth of the accretion column, 
disappearance of the beam 
and enhancement of the lateral X-ray emission illuminating the magnetospheric equator
(see Fig. \ref{f:switch}). Therefore, the radiative energy density 
in the equatorial magnetospheric region strongly increases, Compton cooling
resumes, and the source goes back to higher luminosity levels. 

The idea that the transition between the two regimes may be triggered by a change in the X-ray beam pattern is supported by the pulse profile observations of Vela X-1 
in different energy bands \cite{Doroshenko_ea11}. 
The observed change in phase of the 20--60 keV profile in the off-state (at X-ray luminosity $\sim 2.4\times 10^{35}$~erg~s$^{-1}$), reported by \cite{Doroshenko_ea11}, suggests a disappearance of the fan beam at hard X-ray energies upon the source entering this state and the formation of
a pencil beam
(see \cite{2013MNRAS.428..670S} for more detailed discussion).

Note that the pulse profile phase change associated with X-ray beam switching below some critical luminosity, as observed in Vela X-1, seems to be suggested by an $XMM-Newton$ observation of the SFXT IGR~J11215--5952 (see Fig.~3 in \cite{Sidoli2007}),
corroborating the subsonic accretion regime with radiative plasma cooling at low X-ray luminosities in SFXTs as well, as we shall discuss in the next section.

\subsection{SFXTs}

Supergiant Fast X-ray Transients (SFXTs) are a subclass of 
HMXBs 
associated with early-type supergiant companions \citep{Pellizza2006, Chaty2008, Rahoui2008},
and characterized by sporadic, short and bright X--ray flares 
reaching peak luminosities of 10$^{36}$--10$^{37}$~erg~s$^{-1}$.
Most of them were discovered by INTEGRAL \citep{2003ATel..157....1C, 2003ATel..176....1M, 2003ATel..190....1S, 
2003ATel..192....1G, Sguera2005, Negueruela2005}.
They show high dynamic ranges (between 100 and 10,000, depending on the specific source; 
e.g. \citep{Romano2011, 2014A&A...562A...2R}) and their X-ray spectra in outburst are very similar to accreting pulsars in HMXBs.  
In fact, half of them have measured neutron star (NS) spin periods similar to those observed 
from persistent HMXBs (see \citep{2012int..workE..11S} for a recent review).

The physical mechanism driving their transient behavior, related to the accretion by the compact
object of matter from the supergiant wind, has been discussed by several authors
and is still a matter of debate, as some of them require particular properties of the compact objects hosted in these systems 
\citep{2007AstL...33..149G, 2008ApJ...683.1031B}, 
and others assume
peculiar clumpy properties of the supergiant winds and/or orbital characteristics 
\citep{zand2005,Walter2007,
Sidoli2007,
Negueruela2008,2009MNRAS.398.2152D,Oskinova2012}. 
Recent studies of HMXB population in the Galaxy  \citep{2013MNRAS.431..327L}
suggested that SFXT 
activity should be connected to some accretion inhibition from the stellar wind. 

\textbf{Energy released in bright flares.}
The typical energy released in a SFXT bright flare is about 
$10^{38}-10^{40}$~ergs \cite{2014arXiv1405.5707S}, 
varying by one order of magnitude between different
sources. That is, the mass fallen onto the NS
in a typical bright flare varies from $10^{18}$~g to around $10^{20}$~g. 

The typical X-ray luminosity outside outbursts in SFXTs is about 
$L_{x,low}\simeq 10^{34}$~erg s$^{-1}$ \citep{2008ApJ...687.1230S},
 and below we shall normalise the luminosity to this value, $L_{34}$. 
At these low X-ray luminosities, the plasma entry rate into the magnetosphere is controlled 
by radiative plasma cooling.
Further, it is convenient to normalise the typical stellar wind velocity from hot OB-supergiants $v_w$ 
to 1000~km~s$^{-1}$ (for orbital periods of about a few days or larger the NS orbital velocities can be neglected compared to the stellar wind velocity from the OB-star), so that the Bondi gravitational capture radius is $R_B=2GM/v_w^2=4 \times 10^{10}v_{8}^{-2}$~cm 
for a fiducial NS mass of $M_x=1.5 M_\odot$.

Let us assume that a quasi-static shell hangs over the magnetosphere around the NS, with the magnetospheric accretion rate being controlled by radiative plasma cooling.
We denote the actual steady-state accretion rate as $\dot M_a$ so that the observed X-ray steady-state luminosity  is $L_x=0.1\dot M_a c^2$. Then from the theory of subsonic 
quasi-spherical accretion \citep{2012MNRAS.420..216S} we know that the factor $f(u)$ (the ratio of the actual velocity of plasma entering the magnetosphere, due to the Rayleigh-Taylor instability, 
to the free-fall velocity at the magnetosphere,
$u_{ff}(R_{A})=\sqrt{2GM/R_A}$) reads \citep{2013MNRAS.428..670S,2014EPJWC..6402001S}
\beq{furad}
f(u)_{rad} \simeq 0.036 L_{34}^{2/9}\mu_{30}^{2/27}\,.
\eeq
(See also Eq. (\ref{fu_rad1}) above).

The shell is quasi-static (and likely convective), unless something triggers a 
much faster matter fall through the magnetosphere (a possible reason is suggested below). 
It is straightforward to calculate the mass of the shell 
using the density distribution $\rho(R)\propto R^{-3/2}$ 
\citep{2012MNRAS.420..216S}. Using the mass continuity equation to eliminate the density above the magnetosphere, we readily find 
\beq{deltaM}
\Delta M \approx \frac{2}{3} \frac{\dot M_a}{f(u)}t_{ff}(R_B)\,.
\eeq
Note that this mass can be expressed through measurable quantities
$L_{x,low}$, $\mu_{30}$ and the (not directly observed) stellar wind
velocity at the Bondi radius $v_w(R_B)$. Using Eq. (\ref{furad}) for  
the radiative plasma cooling, we obtain
\beq{deltaMrad}
\Delta M_{rad} 
\approx 8\times 10^{17} [g] L_{34}^{7/9} v_8^{-3}\mu_{30}^{-2/27}\,.
\eeq
The simple estimate (\ref{deltaMrad}) shows that for a typical wind velocity 
near the NS of about 500~km~s$^{-1}$ the \emph{typical} mass of the hot 
magnetospheric shell is around $10^{19}$~g, 
corresponding to $10^{39}$~ergs released in a flare in which all the matter from the shell
is accreted onto the NS, as observed. 
Clearly, variations in stellar wind velocity between different sources by a factor of $\sim 2$  
would produce the one-order-of-magnitude spread in $\Delta M$ observed in bright SFXT flares.

\begin{figure*}
\includegraphics[width=0.7\textwidth]{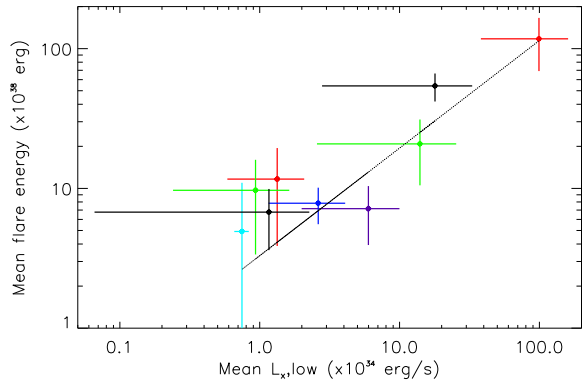}
\caption{The mean energy released in bright flares (17 -- 50 keV, data
from \citep{PaizisSidoli2014}) versus average
\textit{INTEGRAL}/IBIS source luminosity.
The \textit{x-axis}
is in units of $10^{34}$~erg~s$^{-1}$,
the \textit{y-axis} is in units of $10^{38}$~ergs.
The straight line gives the formal rms linear fit with the slope 
$0.77 \pm 0.13$. (Figure adapted from \cite{2014arXiv1405.5707S}). }
\label{fig:dM}
\end{figure*}

In Fig. \ref{fig:dM} we show the mean energy of SFXT bright flares 
$\Delta E=0.1\Delta M c^2$ as a function of the low (non flaring) X-ray luminosity 
for nine SFTXs from our recent paper \cite{2014arXiv1405.5707S}. 
The low (non flaring) X-ray luminosity (\emph{x-axis}) has been taken from 
\cite{krivonos2012}, where a nine year time-averaged source flux in the 17--60\,keV band is given for each source \footnote{IGR\,J17544--2619, IGR\,J16418--4532, IGR\,J16479--4514, IGR\,J16465--4507, SAX\,J1818.6--1703,  IGR\,J18483--0311, XTE\,J1739--302, IGR\,J08408--4503, IGR\,J18450--0435, IGR\,J18410--0535, IGR\,J11215--5952}. The data selection and analysis is discussed in detail in \cite{PaizisSidoli2014}, together with the assumed distances and relevant references, so we refer the reader to that paper for the technical details.
The uncertainties on the low luminosities  
include both the statistical errors on source fluxes, as reported in \cite{krivonos2012},
and the known SFXTs distances and their uncertainties as reported by  \cite{PaizisSidoli2014}. 
The formal rms fit to these points, shown by the straight line, gives the dependence of 
$\Delta E_{38}=(3.3\pm 1.0)L_{34}^{0.77\pm  0.13}$. 
This exactly corresponds to the radiative cooling regime $\Delta E\propto L^{7/9}$ 
(see Eq. (\ref{deltaMrad})), as expected.  
A comparison with the coefficient in expression (\ref{deltaMrad}) suggests $v_8\sim 0.62$,  
similar to typical wind velocities observed in HMXBs.

What can trigger SFXT flaring activity? As noted in 
\citep{2013MNRAS.428..670S},
if there is an instability leading to a rapid fall of matter through the magnetosphere, 
a large quantity of X-ray photons produced near the NS surface should 
rapidly cool down the plasma near the magnetosphere, further increasing the plasma fall velocity
$u_R(R_A)$ 
and the ensuing accretion NS luminosity $L_x$. Therefore, in a bright flare 
the entire shell can fall onto the NS on the free-fall time scale from the outer 
radius of the shell $t_{ff}(R_B)\sim 1000$~s. Clearly, the shell will be replenished by
new wind capture, so the flares will repeat as long as the 
rapid mass entry rate into the magnetosphere is sustained.

\vskip\baselineskip
\noindent
\textbf{Magnetized stellar wind as the flare trigger.}
We suggest that the shell instability described above can be 
triggered by a large-scale magnetic field sporadically 
carried by the stellar wind of the optical OB 
companion. Observations suggest that about $\sim 10\%$ of hot OB-stars have magnetic fields
up to a few kG (see \citep{2013arXiv1312.4755B} for a recent review and discussion).
It is also well known from Solar wind studies (see e.g. reviews \citep{2004PhyU...47R...1Z, lrsp-2013-2} and references therein) that the Solar wind patches carrying tangent magnetic fields 
has a  lower velocity (about $350$~km~s$^{-1}$) than the wind with radial magnetic fields 
(up to $\sim 700$~km s$^{-1}$). Fluctuations of the stellar wind density and velocity 
from massive stars are also known from spectroscopic observations \citep{2008A&ARv..16..209P}, 
with typical velocity fluctuations up to $0.1\ v_\infty\sim 200-300$~km s$^{-1}$.  

The effect of the magnetic field carried by the stellar wind is twofold: first, 
it may trigger
rapid mass entry to the magnetosphere via magnetic reconnection in the magnetopause (the phenomenon well known in the dayside Earth magnetosphere, \citep{1961PhRvL...6...47D}), and secondly, 
the magnetized parts of the wind (magnetized clumps with a tangent magnetic field) have a lower velocity than the non magnetised 
ones (or the ones carrying the radial field). As discussed in \cite{2014arXiv1405.5707S} and below,
magnetic reconnection 
can increase the plasma fall velocity in the shell from inefficient, radiative-cooling controlled settling accretion 
with $f(u)_{rad}\sim 0.03-0.1$, 
up to the maximum possible free-fall velocity with $f(u)=1$.
In other words, during a bright flare subsonic 
settling accretion turns into supersonic Bondi accretion.
The second factor (slower wind velocity in magnetized clumps with tangent magnetic field) 
strongly increases the Bondi radius $R_B\propto v_w^{-2}$
and the corresponding Bondi mass accretion rate $\dot M_B\propto v_w^{-3}$. 

Indeed, we can write down the mass accretion rate onto the NS in the unflaring
(low-luminosity) state as $\dot M_{a,low}=f(u) \dot M_B$
with $f(u)$ given by expression (\ref{furad}) 
and $\dot M_B\simeq \pi R_B^2 \rho_w v_w $.
Eliminating the wind density $\rho_w$ using the mass continuity equation written for the 
spherically symmetric stellar wind from the optical star with power $\dot M_o$ and assuming  
a circular binary orbit, we arrive at 
$
\dot M_B\simeq \frac{1}{4}\dot M_o \myfrac{R_B}{a}^2\,.
$
Next, let us utilize 
the well-known relation for the radiative wind mass-loss rate from massive hot stars
$
\dot M_o\simeq \epsilon \frac{L}{cv_\infty}
$
where $L$ is the optical star luminosity, $v_\infty$ is the stellar wind velocity at infinity,
typically 2000-3000 km s$^{-1}$ for OB stars and $\epsilon\simeq 0.4-1$ is the efficiency factor \citep{1976A&A....49..327L} (in the numerical estimates below we shall assume $\epsilon=0.5$). 
It is also possible to reduce the luminosity $L$ of a
massive star to its mass $M$ using 
the phenomenological relation $(L/L_\odot)\approx 19 (M/M_\odot)^{2.76}$ (see e.g. \citep{2007AstL...33..251V}). Combining the above equations and using
Kepler's third law to express the orbital separation $a$ through the binary period $P_b$, we find for the X-ray luminosity of SFXTs in the non-flaring state 
\begin{eqnarray}
\label{Lxlow}
L_{x,low}\simeq & 5\times 10^{35} [\hbox{erg~s}^{-1}] f(u) 
\myfrac{M}{10 M_\odot}^{2.76-2/3} \nonumber\\
&\myfrac{v_\infty}{1000 \mathrm{km~s}^{-1}}^{-1}
\myfrac{v_w}{500 \mathrm{km~s}^{-1}}^{-4}\myfrac{P_b}{10 \mathrm{d}}^{-4/3}\,,
\end{eqnarray}
which for $f(u)\sim 0.03-0.1$ 
corresponds to the typical low-state luminosities of SFXTs of $\sim 10^{34}$~erg~s$^{-1}$. 

It is straightforward to see that a transition from the low state (subsonic accretion with 
slow magnetospheric entry rate $f(u)\sim 0.03-0.1$) to supersonic free-fall Bondi accretion 
with $f(u)=1$ due to the magnetized stellar wind with the velocity decreasing by a factor of two, for example, would lead to a flaring luminosity of $L_{x,flare}\sim (10\div 30)\times 2^5 L_{x,low}$. This shows that 
the dynamical range of SFXT bright flares ($\sim 300-1000$) can be naturally reproduced by the proposed mechanism.

\textbf{Conditions for magnetic reconnection.}
For magnetic field reconnection to occur, the time the magnetized plasma spends 
near the magnetopause should be at least comparable to 
the reconnection time, $t_r\sim R_A/v_r$, where
$v_r$ is the magnetic reconnection rate, which is difficult to assess from first principles
\citep{2009ARA&A..47..291Z}.
For example, in the Petschek fast reconnection model $v_r=v_A(\pi/8\ln S)$, where $v_A$ is the 
Alfv\'en speed and $S$ is the Lundquist number (the ratio of the global Ohmic dissipation 
time to the Alfv\'en time); for typical conditions near NS magnetospheres we
find $S\sim 10^{28}$ and $v_r\sim 0.006 v_A$. In real astrophysical plasmas 
the large-scale magnetic reconnection rate can be a few times as high, 
$v_r\sim 0.03-0.07 v_A$ \citep{2009ARA&A..47..291Z}, and, guided by phenomenology, we can parametrize it as $v_r=\epsilon_r v_A$ with $\epsilon_r\sim 0.01-0.1$. The longest time-scale the plasma penetrating into the magnetosphere spends near the magnetopause
is the instability time, $t_{inst}\sim t_{ff}(R_A)f(u)_{rad}$ \citep{2012MNRAS.420..216S}, so the 
reconnection may occur if 
$t_r/t_{inst}\sim (u_{ff}/v_A)(f(u)_{rad}/\epsilon_r)\lesssim 1$. As near 
$R_A$ (from its definition) $v_A\sim u_{ff}$, we arrive at $f(u)_{rad}\lesssim\epsilon_r$ as
the necessary reconnection condition. According to Eq. (\ref{furad}), it is satisfied only 
at sufficiently low X-ray luminosities, pertinent to 'quiet' SFXT states. 
\textit{This explains why in HMXBs with convective shells at higher luminosity
(but still lower than $4\times 10^{36}$~erg~s$^{-1}$, at which settling accretion is possible), 
%
reconnection from magnetised plasma accretion will not
lead to shell instability, but only 
to temporal establishment of the 'strong coupling regime' 
of angular momentum transfer through the shell, as 
discussed in \citep{2012MNRAS.420..216S}.} 
Episodic strong spin-ups, as observed in GX 301-2, 
may be manifestations of such 'failed' 
reconnection-induced shell instability.

Therefore, it seems likely that the key difference between 
steady HMXBs like Vela X-1, GX 301-2 (showing only moderate flaring activity) and SFXTs is
that in the first case the effects of possibly magnetized stellar winds from optical OB-companions
are insignificant (basically due to the rather high mean accretion rate),
while in SFXTs with lower 'steady' X-ray luminosity, 
large-scale magnetic fields, sporadically carried by clumps in the wind, 
can trigger SFXT flaring activity via magnetic reconnection near the magnetospheric boundary. 
The observed power-law SFXT flare distributions, discussed in \citep{PaizisSidoli2014},
with respect to the log-normal distributions for classical HMXBs \citep{2010A&A...519A..37F}, 
may be related to the properties of magnetized stellar wind and physics of its interaction 
with the NS magnetosphere.

\section{Conclusions}

In \cite{2012MNRAS.420..216S,2013arXiv1302.0500S,2014EPJWC..6402001S} 
a theory of quasi-spherical wind accretion onto slowly rotating magnetized NS 
in binary systems was developed. It was shown that at luminosities below $\sim 4\times 10^{36}$
erg s$^{-1}$, the accreting plasma does not cool before reaching the magnetosphere,
so a hot quasi-static shell forms around the NS magnetosphere. 
This shell is likely to be convectively unstable and turbulent, which 
allows angular momentum to be transferred to/from the rotating magnetosphere
causing NS spin-up/spin-down with specific behaviour as a function of the  
X-ray luminosity \cite{2012MNRAS.420..216S}. The theory is able to explain 
the observed spin-up/spin-down behavior and phenomenological 
correlations in slowly rotating moderately luminous
X-ray pulsars (Vela X-1, GX 301-2 \cite{2012MNRAS.420..216S}), as well
as the properties of the 
steady spin-down trend observed in GX 1+4 \cite{2012A&A...537A..66G}
and other slowly rotating low-luminosity X-ray pulsars (e.g. SXP 1062, 4U 2206
+54) \cite{2014EPJWC..6402002P} without invoking extremely strong
NS magnetic fields.

In the settling accretion theory, the actual mass accretion rate 
through the shell is determined by the ability of the plasma to enter the NS magnetosphere
via the Rayleigh-Taylor instability, which depends on the cooling mechanism.
As shown in \cite{2013MNRAS.428..670S}, the change between different plasma
cooling regimes (from more efficient Compton cooling to less efficient 
radiative cooling) can occur when the X-ray luminosity decreases below some 
value $L_\dag\sim 3\times 10^{35}$~erg~s$^{-1}$. This transition occurs mainly due to 
a switch of the X-ray beam pattern from a fan-like form generated by the accretion column 
(at high luminositites) to a pencil-like form (at low luminosities). This X-ray diagram switch
is accompanied with an X-ray pulse profile shift, as observed during the 
temporal low-luminosity 'off' states in Vela X-1  \cite{Doroshenko_ea11}, as well
as in other slowly rotating accreting NSs (e.g. in the low states of 
SFXT IGR~J11215--5952 \cite{Sidoli2007}).

The subsonic settling accretion theory can also be 
applied to explain strongly non-stationary phenomena, including SFXT 
flares. As argued in \cite{2014arXiv1405.5707S},  
SFXT bright flares may be caused by sporadic transitions between 
different regimes of accretion in a quasi-spherical
shell around a slowly rotating magnetized neutron star. The non-flaring steady 
states of SFXTs with low X-ray luminosity $L_{x,low}\sim 10^{32}-10^{34}$~erg~s$^{-1}$ may be  
associated with settling subsonic accretion mediated by 
ineffective radiative plasma cooling near the magnetospheric boundary. 
In this state, the accretion rate onto the neutron star is suppressed 
by a factor of $\sim 30$ relative to the Bondi-Hoyle-Littleton value. 
Changes in the local wind velocity and density due to, e.g., clumps, 
can only increase the mass accretion rate
by a factor of $\sim 10-30$, 
bringing the system into the Compton cooling dominated regime, 
and lead to the production of moderately bright flares ($L_x\lesssim 10^{36}$~erg~s$^{-1}$).
To interpret the brightest flares ($L_x>10^{36}$~erg~s$^{-1}$) displayed by the SFXTs within the quasi-spherical settling accretion regime,
a larger increase in the mass accretion rate can be produced by 
sporadic capture of magnetized stellar wind plasma. 
Such episodes should not be associated with specific binary orbital phases, as 
observed in e.g. IGR J17544-2619 \citep{2014MNRAS.439.2175D}. 
At sufficiently low accretion rates, 
magnetic reconnection 
can enhance the magnetospheric plasma entry rate, resulting in copious production of X-ray photons,
strong Compton cooling and ultimately in unstable 
accretion of the entire shell. 
A bright flare develops on the free-fall time
scale in the shell, and the typical energy released in an SFXT bright flare corresponds to 
the entire mass of the shell. This view is consistent with 
the energy released in SFXT bright flares ($\sim 10^{38}-10^{40}$~ergs),
their typical dynamic range ($\sim 100-1000$), and with the observed dependence of these characteristics on the average unflaring X-ray luminosity of SFXTs. 
Thus the flaring behaviour of SFXTs, as opposed to steady HMXBs, can be primarily related to 
their low X-ray luminosity in the settling accretion regime, 
allowing sporadic magnetic reconnection to occur during 
magnetized plasma entry to the NS magnetosphere.

We conclude that the settling regime of accretion onto NSs in wind-accreting 
close binary systems with moderate and low X-ray luminosity 
is now supported by different types of observations. There are likely to be more to come with the ever increasing
quality of X-ray observations of accreting magnetised neutron stars.

\textbf{Acknowledgement.}{The work is supported by the Russian Science Foundation grant 14-12-00146.}

\bibliographystyle{azh}
\bibliography{wind_z}
\end{document}